\documentclass[conference]{IEEEtran}

\usepackage{cite}
\usepackage{color}
\usepackage{balance}
\ifCLASSINFOpdf
\usepackage[pdftex]{graphicx}
\usepackage{epstopdf}
\usepackage{subfigure}
\usepackage{pgfplots,multicol}
\usepackage{amsmath,stmaryrd,pict2e,picture}
\else
\fi
\usepackage{amsmath,amssymb,amsthm}
\usepackage{mathtools}
\usepackage{pifont}

\newlength{\rowidth}
\AtBeginDocument{\setlength{\rowidth}{3em}}
\usepackage{mathrsfs}
\usepackage{algorithm, algorithmic}

\def\footnoterule{\relax%
	\kern-5pt
	\hbox to \columnwidth{\hfill\vrule width 0.75\columnwidth height 0.4pt\hfill}
	\kern4.6pt}
\makeatother
\IEEEoverridecommandlockouts

	\title{Cognitive-Driven Optimization of Sparse Array Transceiver for MIMO Radar Beamforming}
	\author{\IEEEauthorblockN{ Weitong Zhai$^1$, Xiangrong Wang$^1$, Syed~A.~Hamza$^2$ and
			Moeness~G.~Amin$^3$}
		\IEEEauthorblockA{$^1$School of Electronic and Information Engineering, Beihang University, Beijing, 100191, China\\
		$^2$School of Engineering, Widener University, Chester, PA 19013, USA\\
			$^3$Center for Advanced Communications, Villanova University, Villanova, PA 19085, USA\\
			Emails: \{wtzhai, xrwang\}@buaa.edu.cn, shamza@widener.edu, moeness.amin@villanova.edu}
\thanks{The work by W Zhai and X Wang is supported by National Natural Science Foundation of China under Grant No. 62071021 and No. 61827901.}}
\begin{document}
\maketitle
\begin{abstract}
Cognitive multiple-input multiple-output (MIMO) radar is capable of adjusting system parameters adaptively by sensing and learning in complex dynamic environment. Beamforming performance of MIMO radar is guided by both beamforming weight coefficients and the transceiver configuration. We propose a cognitive-driven MIMO array design where both the beamforming weights and the transceiver configuration are adaptively and concurrently optimized under different environmental conditions. The perception-action cycle involves data collection of full virtual array, covariance reconstruction and joint design of the transmit and receive arrays by antenna selection.
The optimal transceiver array design is realized by promoting two-dimensional group sparsity via iteratively minimizing reweighted mixed $l_{2,1}$-norm, with constraints imposed on transceiver antenna spacing for proper transmit/receive isolation. Simulations are provided to demonstrate the ``perception-action'' capability of the proposed cognitive sparse MIMO array in achieving enhanced beamforming and anti-jamming in dynamic \textcolor{black}{target and interference} environment.

\end{abstract}

\IEEEpeerreviewmaketitle
\section{Introduction}
\begin{figure}[!t]
	\centering
	\includegraphics[height=2in, width=3.5in]{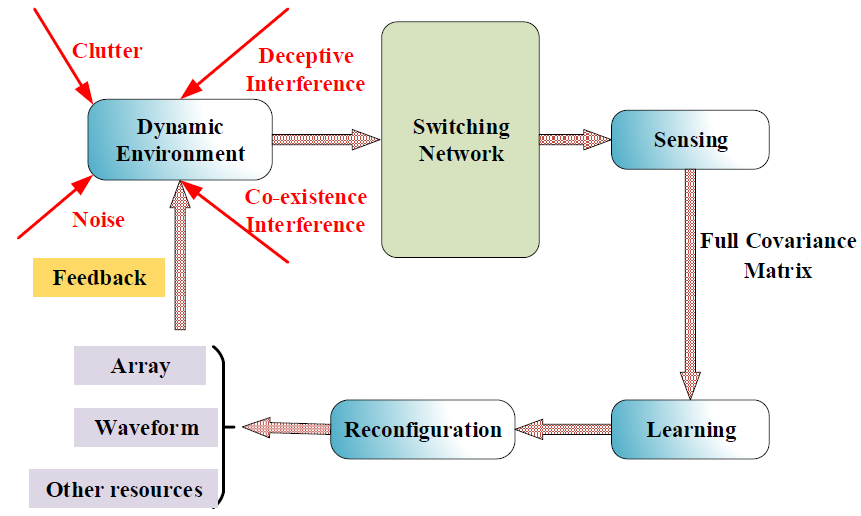}
	\caption{Schematic diagram of cognitive MIMO Radar.}
	\vspace{-0.3cm}
	\label{Fig.1}
\end{figure}

Multiple-Input-Multiple-Output (MIMO) radars enable superior capabilities compared with standard phased arrays radar. Cognitive radar continuously interacts with the environment and update the radar parameters through the acquired knowledge \cite{S.Haykin,8398580}. In MIMO radar, ``perception-action'' cognition approach typically entails adaptive optimizations of the transmit waveforms and power allocation \cite{6541985,7131266,7185382}. In addition to these parameters, array reconfiguration (hardware) can significantly improve system performance beyond that achieved with fixed antenna positions \cite{1374907, 7444122}. Motivated by this fact, we propose a cognitive-driven sparse MIMO array design method that incorporates sparse transceiver array optimization. The schematic system diagram is shown in Fig. \ref{Fig.1}, where a fully-connected switching network is employed for cognitive array reconfiguration. The task is to continuously and cognitively select different subsets of antennas from a large array to deliver the best performance under time-varying environment. 

Sparse array design, aided by emerging fast antenna switching technologies, can lower the overall system cost by reducing the number of expensive front-end processing channels. The conventional MIMO radar has a sparse transceiver array. The inter-element spacing of the receive array is half wavelength, whereas that of the transmit array is multiple wavelengths. Hence, the MIMO sum coarray is a compact uniform linear array (ULA) with a large virtual aperture that enables high spatial resolution \cite{7962219,7131164,5728938}. This configuration, however, may not render an optimum beamforming in terms of maximizing the output signal-to-interference-plus-noise ratio (MaxSINR) \cite{5989873,8438940}. Sparse MIMO array beamforming design can achieve optimization of the transmit sensor locations and the corresponding waveform correlation matrix for synthesizing a desired beampattern \cite{6951471,8681265,6650099}. This approach, however, essentially pursues an decoupled transmit/receive design. In this paper, we pursue the coupled design approach and seek optimum sparse transceiver, configuring the MaxSINR beamformer at the MIMO radar receiver. We assume a colocated MIMO radar platform with orthogonal transmit waveforms. The adaptive beamforming is implemented on the virtual array after matched filtering. In the proposed design, the transceiver array configuration and beamforming weights are concurrently optimized to seek the best output SINR performance in a dynamic target and interference environment. This requires a cognitive operation where the present sensed system information is utilized to determine the next system parameters.

In the first step of the proposed cognitive beamforming, the MIMO radar senses the environment by sequentially switching to different sets of antennas for data collection, based on which a full covariance matrix of the large virtual array is constructed. In the second step of learning, a constrained design optimization problem is formulated and solved to simultaneously provide the configurations of both the transmit and receive arrays associated with MaxSINR. The radar will then act by switching on the selected transmit and receive antennas, and applying the resultant optimum beamforming weights after matched filtering. It is noted that cycle is triggered each time the output SINR degrades from a certain threshold performance, thereby making antenna selection a cognitive operation. A reweighted mixed $l_{2,1}$-norm minimization is employed  to promote a two-dimensional group sparsity for MIMO transceiver design. Moreover, for practical consideration, the optimum cognitive MIMO radar is designed under antenna isolation  constraints to control the minimum spacing between transmitter and receiver antennas.

The rest of this paper is organized as follows: In the section \ref{sec:sensing} of sensing, we construct the covariance matrix of the full virtual array by sequentially switching among different sets of antennas. In the section \ref{sec:learning} of learning, we elaborate on the optimal sparse transceiver design by promoting two-dimensional group sparsity and taking practical constraints into account. In the section \ref{rec:reconfiguraion} of reconfiguration, we reconfigure the beamformer of MIMO radar based on the results of the first two steps. In the following section, we demonstrate the superiority of the proposed optimal transceiver sparse array by simulations. The paper ends with concluding remarks.

\section{Full covariance construction}
\label{sec:sensing}

In the first step, the MIMO radar senses the RF environment by sequentially switching to different sets of sparse arrays. This action enables data collection and a follow-on construction of the full covariance matrix corresponding to the virtual transceiver after matched filtering. Let us first consider a cognitive MIMO radar with a total of $M$ antennas. Later, only $N$ transmit antennas and $N$ receive antennas will be optimally selected from the $M$ available antennas.  The full covariance matrix corresponding to co-located $M$-antenna transmit and receive arrays is constructed as follows.

The received signal at the $n$th time instant is given by,
\begin {equation} 
\label{xn}
\begin{split}
\mathbf{x}(n)=&\eta_{0}(n)\mathbf{a}_{r}(\theta_{0})\mathbf{a}_{t}^{T}(\theta_{0})\mathbf{s}(n)+\sum_{l1=1}^{L_{1}} j_{l_{1}}(n) \mathbf{a}_{r}(\theta_{j,l_{1}})\\
&+\sum_{l2=1}^{L_{2}}\eta_{l_{2}}(n)\mathbf{a}_{r}(\theta_{j,l_{2}})\mathbf{a}_{t}^{T}(\theta_{j,l_{2}})\mathbf{s}(n)+\mathbf{v}(n),
\end{split}
\end{equation}
where $\mathbf{s}(n)\in \mathbb{C}^{M\times 1}$ represents the $M$ orthogonal transmit signals, $\theta_{0}$ denotes the target angle, $j_{l_{1}}(n)$ represents the $l_{1}$-th co-existing interference whose arrival angle relative to the receiver is $\theta_{j,l_{1}}$. The variables $\eta_{0}(n)$ and $\eta_{l_{2}}(n)$ represent the reflection coefficients of the target and the $l_{2}$-th deceptive interference, or spoofer, respectively, which are assumed to be uncorrelated and follow complex Gaussian distribution. Deceptive interference is used by adversaries to mimic target echo to interfere with target detection and thus exhibits the same waveform with radar transmit signal. The arrival angle of the $l_{2}$th deceptive interference is $\theta_{j,l_{2}}$,  $\mathbf{v}(n)\in \mathbb{C}^{M\times 1}$ is additive white Gaussian noise. Also, $\mathbf{a}_{t}(\theta)$ and $\mathbf{a}_{r}(\theta)$ represent the steering vectors of the transmit and receive arrays, respectively. As both transmit and receive arrays are selected from the same $M$-antenna uniform array, $\mathbf{a}_{t}(\theta)$ and $\mathbf{a}_{r}(\theta)$ have the same expression and are given by,
%
{\setlength\abovedisplayskip{10pt}
\setlength\belowdisplayskip{-10pt}
  \begin{equation}
  \label{atr}
 \mathbf{a}_{t/r}(\theta)= {}[1 \,  \,  \, e^{j 2 \pi (d/\lambda) cos\theta}  {}\\
 \,  . \,   . \,  . \, \,  \,  \, \,  \,  \, e^{j 2 \pi (M-1) (d/\lambda) cos\theta}]^T.
 \end{equation}}
 
After matched filtering, we obtain the  data matrix $\mathbf{Y}(m) \in \mathbb{C}^{M\times M}$ for the $m$th radar pulse. 
{\setlength\abovedisplayskip{5pt}
\setlength\belowdisplayskip{10pt}
\begin {equation} \label{Ym}
\begin{split}
\mathbf{Y}(m)=&\eta_{0}(m)\mathbf{a}_{r}(\theta_{0})\mathbf{a}_{t}^{T}(\theta_{0})\mathbf{R}+\sum_{l1=1}^{L_{1}}\mathbf{a}_{r}(\theta_{j,l_{1}})\mathbf{i}_{l_{1}}^{T}(m)\\
&+\sum_{l2=1}^{L_{2}}\eta_{l_{2}}(m)\mathbf{a}_{r}(\theta_{j,l_{2}})\mathbf{a}_{t}^{T}(\theta_{j,l_{2}})\mathbf{R}+\mathbf{V}'(m),
\end{split}
\end {equation}}where $\mathbf{R}=\sigma_{s}^{2}\mathbf{I}$ is the auto-correlation matrix of $M$ orthogonal transmit signals with equal power of $\sigma_{s}^{2}$, $\mathbf{i}_{l_{1}}(m)=\sum_n j_{l_1}(n) \mathbf{s}^*(n)$ is the output of matched filtering of the $l_{1}$th interference. Since the interference is not related to the transmit waveforms, $\mathbf{i}_{l_{1}}(m)$ still follows Gaussian distribution. We assume that the reflection coefficients obey the Swerling II target model, i.e, they change from one pulse repetition period to another. Vectorizing the data matrix $\mathbf{Y}(m)$, we generate the data vector received by the virtual array,
{\setlength\abovedisplayskip{10pt}
\setlength\belowdisplayskip{-5pt}
\begin {equation} \label{ym}
\begin{split}
\mathbf{y}(m)=&\text{vec}(\mathbf{Y(m)})\\
=&\eta_{0}(m)\sigma_{s}^{2}\mathbf{b}(\theta_{0})+\sum_{l1=1}^{L_{1}}\mathbf{a}_{r}(\theta_{j,l_{1}})\otimes \mathbf{i}_{l_{1}}(m)\\
&+\sum_{l2=1}^{L_{2}}\eta_{l_{2}}(m)\sigma_{s}^{2}\mathbf{b}(\theta_{j,l_{2}})+\mathbf{v}'(m)
\end{split}
\end {equation}}

For convenience, we denote $\mathbf{b}(\theta_{i})=\mathbf{a}_{r}(\theta_{i})\otimes \mathbf{a}_{t}(\theta_{i})$. The covariance matrix of the full virtual array is given by,
{\setlength\abovedisplayskip{5pt}
\setlength\belowdisplayskip{5pt}
\begin {equation} \label{Rx}
\begin{split}
\mathbf{R}_{x}=E\left \{ \mathbf{y}(m)\mathbf{y}^{H}(m) \right \} \approx \frac{1}{T}\sum_{m=1}^T \mathbf{y}(m)\mathbf{y}^{H}(m)
\end{split}
\end {equation}}where a total of $T$ pulses is assumed for sensing. From Eqs. (\ref{Ym}), (\ref{ym}) and (\ref{Rx}), we can observe that in order to obtain the covariance matrix of the full virtual array, we need to find the data matrix $\mathbf{Y}(m)$.

As the data matrix $\mathbf{Y}$ corresponds to the full virtual array, it is a challenging problem to construct $\mathbf{Y}$ using a sparse array with a small number of antennas. In the sequel, we propose a time-multiplexing method to sequentially switch on different sets of antennas for data collection and followed full covariance construction. Matrix $\mathbf{Y}$ is a Hankel matrix, that is, the elements along each anti sub-diagonal are equal. Thereby, we only need to estimate the elements on the first row and the last column to infer all elements of the full matrix. 

\begin{figure}[!t]
	\setlength{\abovecaptionskip}{0pt}
	\setlength{\belowcaptionskip}{0pt}
	\centering
	\includegraphics[height=0.8in, width=3.5in]{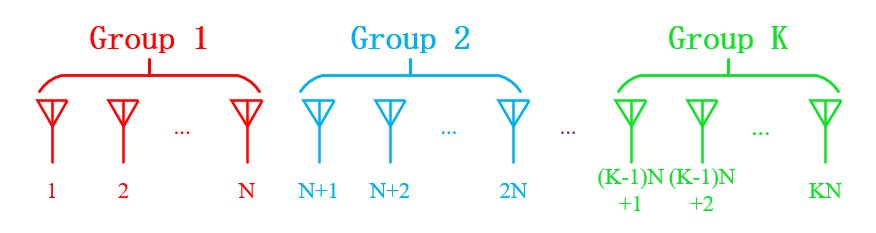}
	\caption{Division of the large $M$-antenna uniform linear array into $K$ groups.}
	\setlength{\abovecaptionskip}{0cm}
	\vspace{-0.6cm}
	\label{Fig.2}
\end{figure}

Suppose we select $N$ transmitters and $N$ receivers from the $M$-antenna array, respectively. In this case, the M-antenna array is divided into $K$ subarrays and each subarray comprises $N$ consecutive antennas as shown in Fig. \ref{Fig.2}; thus $M = KN$. Accordingly, we divide the matrix $\mathbf{Y}$ into $K^{2}$ sub-matrices with the same size of $N\times N$ as follows,
{\setlength\abovedisplayskip{10pt}
\setlength\belowdisplayskip{10pt}
\begin {equation} \label{Y}
\begin{split}
\mathbf{Y}=\begin{bmatrix}
\textcolor{red}{\mathbf{Y}_{11}} & \textcolor{blue}{\mathbf{Y}_{12}} & \textcolor{blue}{...}  & \textcolor{blue}{\mathbf{Y}_{1K}}\\ 
\mathbf{Y}_{21} & \mathbf{Y}_{22} & ... &\textcolor{green}{\mathbf{Y}_{2K}}\\ 
... & ... & ... &\textcolor{green}{...}\\
\mathbf{Y}_{(K-1)1} & \mathbf{Y}_{(K-1)2} & ...  &\textcolor{green}{\mathbf{Y}_{(K-1)K}}\\ 
\mathbf{Y}_{K1} & \mathbf{Y}_{K2} & ... &\textcolor{red}{\mathbf{Y}_{KK}}
\end{bmatrix}
\end{split}
\end {equation}}

Utilizing the Hankel matrix structure, if the $2K-1$ submatrices on the first row and the $K$th column are estimated, as highlighted in color in Eq. (\ref{Y}), we can then recover the entire matrix $\mathbf{Y}$, and subsequently  obtain the covariance matrix using Eqs. (\ref{ym}) and (\ref{Rx}). The $2K-1$ submatrices can be estimated via multiplexing to different sets of antennas according to the following procedure within one radar pulse. 

\begin{itemize}
    \item For submatrix $\mathbf{Y}_{1i}(2\leq i\leq K)$, highlighted in blue, we use the first $N$ antennas as the receiving array, and use the $i$th group containing from $[(i-1)N+1]$th to $(iN)$th antennas to transmit N orthogonal waveforms. After matched filtering at the receiver, we can estimate matrix $\mathbf{Y}_{1i}$, and the other matrices in the first row by sequentially multiplexing to next transmit array.
    \item For submatrix $\mathbf{Y}_{jK}(2\leq j\leq K-1)$, highlighted in green, we use the $[(K-1)N+1]$th to $(KN)$th antennas as the transmit array, and switch the receive array sequentially from $(j-1)N+1$ to $jN$. After matched filtering at the receiver, we can estimate matrix $\mathbf{Y}_{jK}$, and the other matrices in the $K$th column.
 
    \item For submatrix $\mathbf{Y}_{11}$, highlighted in red, we take the 1st to $N$th antennas as the receiving array and the $N+1$th to $(2N)$th antennas as the transmitting array. The set of orthogonal transmit waveforms are then phase rotated by $e^{j(2\pi/\lambda)(Nd)cos\theta_{0}}$, which is equivalent to the selection of the transmit array composed of the 1st to $N$th antennas. After matched filtering at the receiver, we can estimate matrix $\mathbf{Y}_{11}$. The matrix $\mathbf{Y}_{KK}$ can be estimated in the same way.
    
\end{itemize}






\section{Optimal transceiver design}
\label{sec:learning}

After obtaining the covariance matrix of the full virtual array, the next step for cognitive MIMO radar is to determine the optimal transceiver, including both the beamforming weights and array configurations.

\subsection{Beamforming for MIMO Radar}

When the $M$ antennas are used for both the transmit and receive arrays, as the case in Section \ref{sec:sensing}, the output of the MIMO receiver is given by,
 \begin {equation}  
 \label{zm}
 z(m) = \mathbf{w}^H \mathbf{y}(m).
 \end {equation}
where beamforming weight vector $\mathbf{w} \in \mathbb{C}^{M^2 \times 1}$ is applied to the virtual array at the receiver after matched filtering. The optimal of MaxSINR beamformer is referred to as Capon beamfomrer, which can be obtained by minimizing the noise and interference power without weakening the desired signal \cite{8378759,1223538,8892512}. That is,
%
{\setlength\abovedisplayskip{5pt}
\setlength\belowdisplayskip{5pt}
\begin{equation} \label{wHRxw}
\begin{aligned}
\underset{\mathbf{w} \in \mathbb{C}^{M^{2}\times 1}}{\text{minimize}} & \quad   \mathbf{w}^H\mathbf{R_{x}}\mathbf{w},\\
\text{s.t.} & \quad     \mathbf{w}^H\mathbf{b}(\theta_{0}) = 1 , 
\end{aligned}
\end{equation}}where \text{s.t.} represents ``subject to''. The solution of the above constrained optimization problem is given by,
{\setlength\abovedisplayskip{10pt}
\setlength\belowdisplayskip{-5pt}
\begin{equation} \label{wo}
\begin{aligned}
\mathbf{w}_{o}=[\mathbf{b}^{H}(\theta_{0})\mathbf{R}_{x}^{-1}\mathbf{b}(\theta_{0})] ^{-1}\mathbf{R}_{x}^{-1}\mathbf{b}(\theta_{0})
\end{aligned}
\end{equation}} 

\subsection{Sparse Transceiver Design}

\begin{figure}[!t]
	\setlength{\abovecaptionskip}{0pt}
	\setlength{\belowcaptionskip}{0pt}
	\centering
	\includegraphics[height=1.8in, width=3.3in]{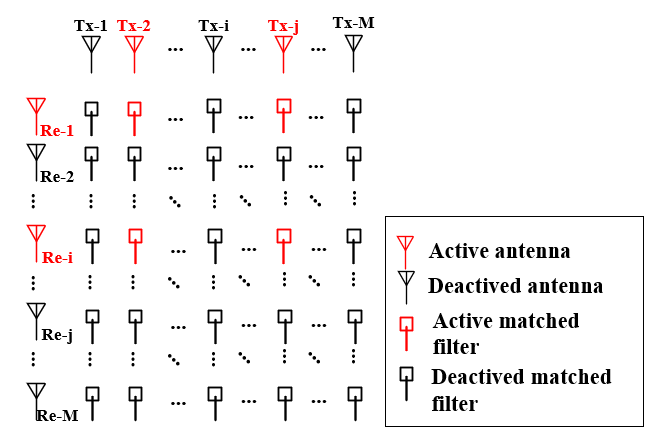}
	\caption{Schematic diagram of virtual sensor for cognitive MIMO radar and the illustration of two-dimensional group sparsity.}
	\label{Fig.3}
    \vspace{-0.5cm}
\end{figure}

For MIMO radar, beamforming in the receiver is implemented on the $M^{2}$-sensor virtual array after matched filtering, as shown in Fig. \ref{Fig.3}. There are $M$ virtual sensors associated with each transmit or receive physical sensor. If a sensor is not activated, it implies that none of the $M$ virtual sensors associated with this sensor are selected. When we leverage the sparsity of beamforming weight vector to design sparse arrays, it suggests that the $M$ consecutive weights vertically or horizontally are all zeros. For example, in Fig. \ref{Fig.3}, the first sensor is not activated for waveform transmission, so none of the virtual sensors in the first column are selected. Similarly, the second sensor is not activated for receiving signals, so all the virtual sensors in the first row are discarded.

It can be inferred from the above analysis that a two-dimensional group sparsity promoting method can be employed to design the sparse transceiver configuration of MIMO radar. Suppose we choose $N$ out of the $M$ sensors as the transmit array and another $N$ different sensors as the receiving array. When a sensor is activated, there is at least one non-zero weight coefficient related to the corresponding $M$ virtual sensors. On the contrary, when a sensor is not activated, the weights of the corresponding $M$ virtual sensors all assume zero values. Thus, the sensor selection scheme needs to warrant that the weights of the entire row or column of virtual sensors in Fig. \ref{Fig.2} are either all zero or have at least one non-zero entry.

The beamforming weight vectors are generally complex valued, whereas the quadratic function in Eq. (\ref{wHRxw}) are real. This observation allows expressing the problem with only real variables, which is typically accomplished by converting the problem from the complex domain to the real domain and concatenating the vectors accordingly,
{\setlength\abovedisplayskip{5pt}
\setlength\belowdisplayskip{10pt}
\begin{multline} \label{k2}
\begin{aligned}
&\tilde{\mathbf{R}}_x=\begin{bmatrix}
\text{real}(\mathbf{R}_x)       & -\text{imag}(\mathbf{R}_x)   \\
\text{imag}(\mathbf{R}_x)       & \text{real}(\mathbf{R}_x)   \\
\end{bmatrix},
\tilde{\mathbf{w}}=\begin{bmatrix}
\text{real}({\mathbf{w}})       \\
\text{imag}({\mathbf{w}})   \\
\end{bmatrix},\\
&\tilde{\mathbf{b}}(\theta_{0})=\begin{bmatrix}
\text{real}(\mathbf{b}(\theta_{0}))       \\
\text{imag}(\mathbf{b}(\theta_{0}))   \\
\end{bmatrix},
\end{aligned}
\end{multline}}where $\tilde{\mathbf{R}}_{x}\in \mathbb{R}^{2M^{2}\times 2M^{2}}$, $\tilde{\mathbf{w}}\in \mathbb{R}^{2M^2\times 1}$ and $\tilde{\mathbf{b}}(\theta_{0})\in \mathbb{R}^{2M^{2}\times 1}$ are the real domain vectors of $\mathbf{R}_{x}$, $\mathbf{w}$ and $\mathbf{b}(\theta_{0})$, respectively.
Therefore, the sensor selection problem of sparse array construction can be described as the following optimization,
{\setlength\abovedisplayskip{5pt}
\setlength\belowdisplayskip{5pt}
\begin{align} 
\underset{\mathbf{\tilde{w}},\mathbf{c},\mathbf{r}}{\text{minimize}} & \quad   \tilde{\mathbf{w}}\tilde{\mathbf{R}}_{x}\tilde{\mathbf{w}} \label{real}\\
\text{s.t.} 
&\quad \tilde{\mathbf{w}}^{H}\tilde{\mathbf{b}}(\theta_{0})=1, \tag{\ref{real}{a}} \label{reala}\\
& \quad ||\mathbf{P}_{i}\odot  \mathbf{\tilde{w}}||_{2}= c_{i},\quad i=1,...,M \tag{\ref{real}{b}} \label{realb}\\
& \quad ||\mathbf{c}||_{0}= N, \tag{\ref{real}{c}} \label{realc}\\
& \quad ||\mathbf{Q}_{j}\odot  \mathbf{\tilde{w}}||_{2}= r_{j},\quad j=1,...,M \tag{\ref{real}{d}} \label{reald}\\
& \quad ||\mathbf{r}||_{0}= N, \tag{\ref{real}{e}} \label{reale}\\
& \quad \mathbf{c}^{H}\mathbf{r}=0, \tag{\ref{real}{f}} \label{realf}
\end{align}}
and 
{\setlength\abovedisplayskip{5pt}
\setlength\belowdisplayskip{5pt}
\begin{align} \label{Pi}
\mathbf{P}_{i}=[\hspace{-3mm}\overbrace{0  ...0...0}^{\mbox{\footnotesize$\begin{array}{c}M\ elements\\
1st\ group \end{array}$}}\hspace{-3mm} ...\hspace{-3mm} \overbrace{1...1...1}^{\mbox{\footnotesize$\begin{array}{c}M\ elements\\
 ith\ group \end{array}$}} \hspace{-3mm}...\hspace{-3mm} 
 \overbrace{1...1...1}^{\mbox{\footnotesize$\begin{array}{c}M\ elements\\
 (M+i)th\ group \end{array}$}} \hspace{-2mm}...\hspace{-3mm}\overbrace{0...0...0}^{\mbox{\footnotesize$\begin{array}{c}M\ elements\\
    2Mth\ group \end{array}$}}\hspace{-3mm}]^{H},
\end{align}}
{\setlength\abovedisplayskip{5pt}
\setlength\belowdisplayskip{10pt}
\begin{align} \label{Qj}
\mathbf{Q}_{j}=[\ \overbrace{\underbrace{0\  ...\ 0}_{(j-1)\ 0s}\ 1\ 0\ ...\ 0}^{\mbox{\footnotesize$\begin{array}{c}
M\ elements\ of\\
the\ 1st\ group \end{array}$}}\ .\ .\ . \ \overbrace{\underbrace{0\  ...\ 0}_{(j-1)\ 0s}\ 1\ 0\ ...\ 0}^{\mbox{\footnotesize$\begin{array}{c}
M\ elements\ of\\
    the\ 2Mth\ group \end{array}$}}]^{H}.
\end{align}}where $\odot$ denotes the element-wise product and $\mathbf{P}_{i}$ is the transmission selection matrix, which is used to select the real and imaginary parts of the weight associated with the $i$th transmitting sensor. Similarly, $\mathbf{Q}_{j}$ is the receiving selection matrix, which is used to select the real and imaginary parts of the weight associated with the $j$th receiving sensor. Constraints (\ref{realc}) and (\ref{reale}) indicate that N sensors are selected for each transmitting and receiving array, and $||.||_{0}$ denotes the $l_{0}$ norm. Constraint (\ref{realf}) implies that the transmitting array and receiving array do not share any sensors.

\subsection{Reweighted $l_{2,1}$-norm}

The $l_{0}$-norm constraints in (\ref{realc}) and (\ref{reale}) are non-convex, which renders the above optimization problem difficult to solve. We utilize the $l_{1}$-norm to approximate the $l_{0}$-norm \cite{4522554}. To further promote sparsity, an iterative reweighted $l_{1}$-norm was proposed in \cite{Emmanuel2007Enhancing}. Based on this idea, we employ an iterative reweighted mixed $l_{2,1}$-norm to promote two-dimensional group sparsity, which can be described as follows,
{\setlength\abovedisplayskip{2pt}
\setlength\belowdisplayskip{2pt}
\begin{align} \label{iteration}
\underset{\mathbf{\tilde{w}}, \mathbf{c},\mathbf{r}}{\text{minimize}} & \quad   \tilde{\mathbf{w}}\tilde{\mathbf{R}}_{x}\tilde{\mathbf{w}}  +   \alpha(\mathbf{p}^{H}\mathbf{c})  +\beta(\mathbf{q}^{H}\mathbf{r}) \\
\text{s.t.} 
& \quad \tilde{\mathbf{w}}^{H}\tilde{\mathbf{b}}(\theta_{0})=1, \tag{\ref{iteration}{a}}\label{iterationa}\\
& \quad ||\mathbf{P}_{i}\odot  \mathbf{\tilde{w}}||_{2}\leq c_{i}, \tag{\ref{iteration}{b}}\label{iterationb}\\
& \quad 0\leq c_{i}\leq 1,\quad i=1,...,M \tag{\ref{iteration}{c}}\label{iterationc}\\
& \quad ||\mathbf{Q}_{j}\odot \mathbf{\tilde{w}}||_{2}\leq r_{j},\tag{\ref{iteration}{d}}\label{iterationd}\\
& \quad 0\leq r_{j}\leq 1,\quad j=1,...,M \tag{\ref{iteration}{e}}\label{iteratione}\\
& \quad \mathbf{1}_{M}^{H}\mathbf{c}=N, \quad \mathbf{1}_{M}^{H}\mathbf{r}=N, \tag{\ref{iteration}{g}}\label{iterationg}\\
& \quad \mathbf{c}^{H}\mathbf{r}=0, \tag{\ref{iteration}{h}}\label{iterationh}
\end{align}}where $\alpha$ and $\beta$ are two parameters to balance between the array sparsity and output SINR of the transceiver. A common method of updating the reweighting coefficient is to take the reciprocal of the absolute weight value. However, this fails to control the number of elements to be selected. Thus, similar to \cite{2020Sparse}, in order to control the number of selected antennas, we update the reweighting coefficient $\mathbf{p}$ and $\mathbf{q}$ of the $(K+1)$th iteration using the following formula,
{\setlength\abovedisplayskip{2pt}
	\setlength\belowdisplayskip{2pt}
\begin{align}\label{piqj}
{p}_i^{(k+1)}=&\frac{1-c_i^{(k)}}{1-e^{-\beta_{0}c_i^{(k)}}+\epsilon}-(\frac{1}{\epsilon})(c_i^{(k)})^{\alpha_{0}}, \nonumber\\
{q}_j^{(k+1)}=&\frac{1-r_j^{(k)}}{1-e^{-\beta_{0}r_j^{(k)}}+\epsilon}-(\frac{1}{\epsilon})(r_j^{(k)})^{\alpha_{0}},
\end{align}}where $\alpha_{0}$ and $\beta_{0}$ are the parameters controlling the shape of the reweighting curve, and $\epsilon$ prevents the denominator from tending to 0. It can be seen from reference \cite{2020Sparse} that the reweighting coefficient $p_{i}$ can make the value of $c_{i}$ tend to 0 or 1 by setting reward at 1 and imposing punishment at 0. This allows the number of transmitting sensors to be controlled. The same process can be applied to $\mathbf{q}$ for the receiving sensors.

In practical implementation, a close distance between the transmit and receiving antennas will cause power leakage and unwanted coupling effect. To achieve high isolation between the transmit and receive channels, which is always desirable, the selected transmit and receive antennas should be sufficiently spatially separated. Therefore, for the transceiver array design, the two positions neighbouring the selected transmitting antenna should be vacant and not hosting any receive antenna. To satisfy this requirement, we replace the constraint in (\ref{iterationh}) with the following new constraints,
{\setlength\abovedisplayskip{2pt}
	\setlength\belowdisplayskip{0pt}
\begin{equation} \label{g2}
\begin{split}
&c_{i}+r_{i-1}+r_{i}+r_{i+1}\leq 1,\quad i=2,...,M-1,\\
&c_{1}+r_{1}+r_{2}\leq 1,\\
&c_{M}+r_{M}+r_{M-1}\leq 1,
\end{split}
\end{equation}}

\section{new transceiver reconfiguration}
\label{rec:reconfiguraion}

We obtain the optimal transceiver configuration in the current environment through sensing and learning. In the action phase, and according to the two vectors $\mathbf{c}$ and $\mathbf{r}$ from Eq. (\ref{iteration}), we switch on the selected $N$ antennas indicated by $\mathbf{c}$ to transmit the $N$ orthogonal waveforms, and switch on the selected $N$ receiving antennas indicated by $\mathbf{r}$ to receive data. After matched filtering, we perform beamforming with the weights $\mathbf{w}$ obtained from Eq. (\ref{iteration}). Consequently, a new transceiver array will be reconfigured to adapt to the new environment for performance improvement. When the environment changes, the output SINR decreases, instigating the beginning of another ``perception-action'' cycle. 

\section{Simulations} 
\label{Simulations}
In this section, we demonstrate the effectiveness of the proposed cognitive-driven sparse transceiver design for MIMO radar in different scenarios.
\subsection{Example 1}
\begin{figure}[!t]
 	\setlength{\abovecaptionskip}{-5pt}
 	\setlength{\belowcaptionskip}{0pt}
	\centering
	\includegraphics[height=2.2in, width=3.1in]{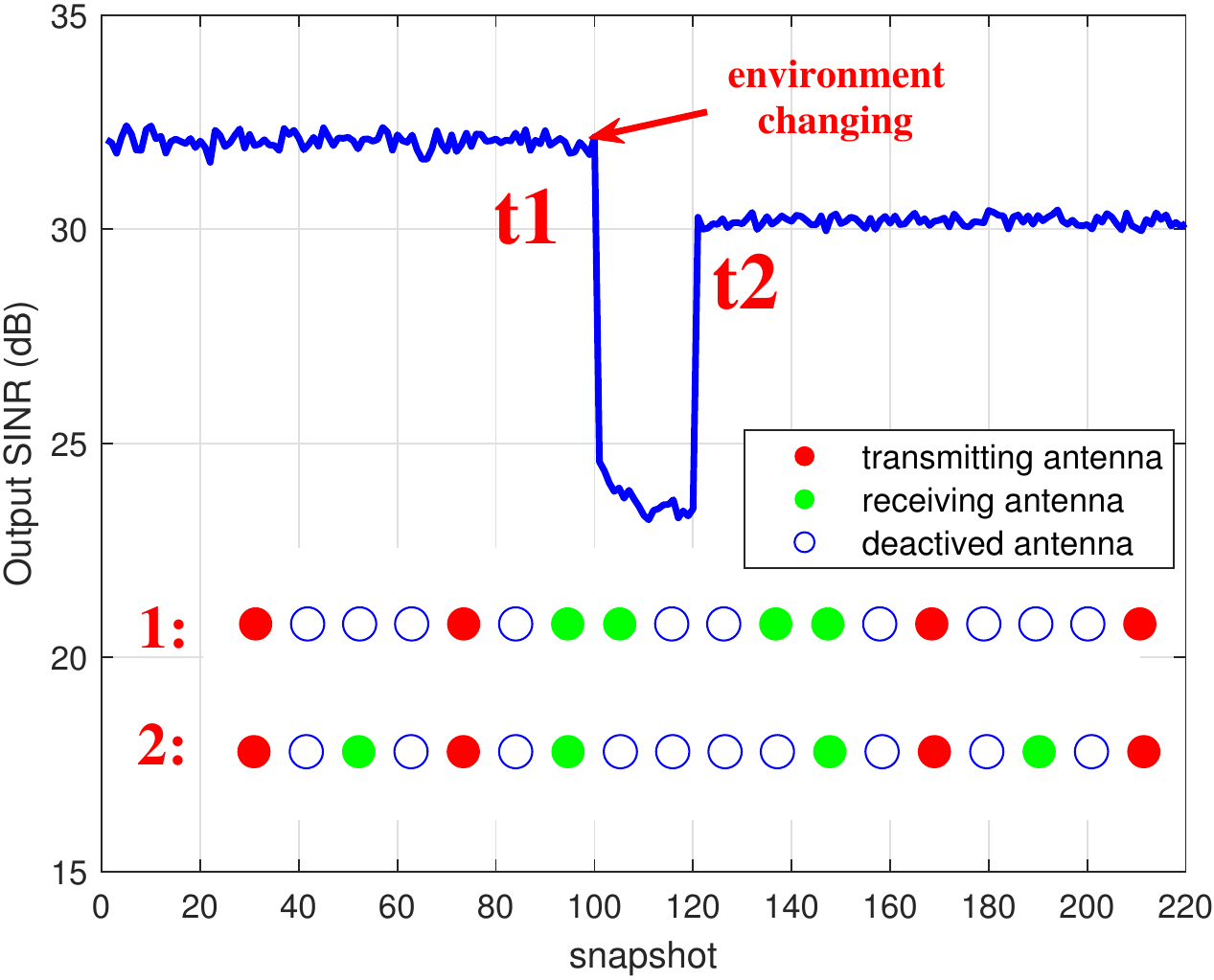}
	\caption{Response of cognitive MIMO radar to environmental changes.}
	\vspace{-0.5cm}
	\label{Fig.4}
\end{figure}

In this example, we simulate the work flow of the proposed cognitive MIMO radar and evaluate its performance. We assume that there are a total of 18 antennas with an inter-element spacing of $\lambda/2$, from which $N=4$ antennas are selected for transmitting and another $N=4$ antennas for receiving. In the first part of the simulation, that is, before time $t_1$, the target is at angle $65^{\circ}$ with the signal-to-noise ratio (SNR) is 20dB. There are two deceptive interferences with an interference-to-noise ratio (INR) of 15dB located at angle $50^{\circ}$ and $60^{\circ}$, respectively. At a certain moment $t_1$, the environmental condition suddenly changes, and another interference close to the target at $63^{\circ}$ switches on. We can observe that the output SINR decreases abruptly starting from time instant $t_1$, as shown in Fig. \ref{Fig.4}. This triggers the cognitive mechanism of the MIMO radar, which re-initiates sensing and then learning from time $t_1$ to time $t_2$. It reconfigures the optimal transceiver arrays responding to the environmental change at time $t_2$. Accordingly, the output SINR is now maximized under the new environment. The two optimum transceiver configurations 1 and 2 before and after the environmental change are shown in Fig. \ref{Fig.4}. Evidently, the selection of transmit antennas remains the same, while the configuration of receiving array has significantly changed. In this experiment, the output SINR of the new environment is slightly lower than that of the initial environment, which is due to the additional interference closed to the expected signal.
 
 \subsection{Example 2}
 
\begin{figure}[!t]
	\setlength{\abovecaptionskip}{-5pt}
	\setlength{\belowcaptionskip}{0pt}
	\centering
	\includegraphics[height=2.2in, width=3.1in]{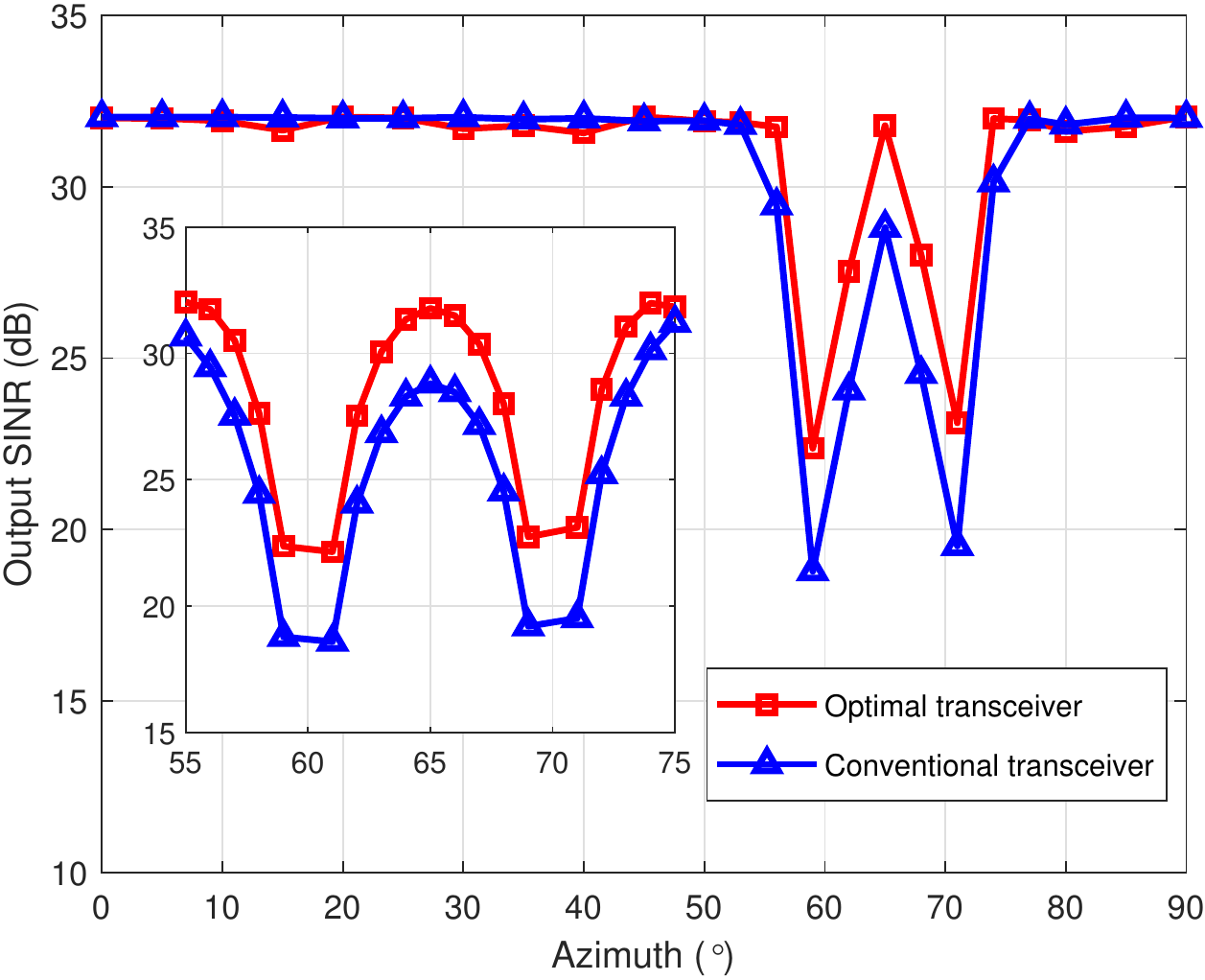}
	\caption{Relationship between output SINR and target angle with the directions of interferences being fixed.}
	\label{Fig.5}
	\vspace{0cm}
\end{figure}

In this example, we compare the performance of the optimal transceiver with that of conventional transceiver that comprises antennas indexed by 1, 5, 9 and 13 for transmitting and antennas indexed by 15 16, 17 and 18 for receiving. Two deceptive interferences are impinging on the array from $60^{\circ}$ and $70^{\circ}$, respectively, with an INR of 15dB. The scanning angle is changing from $0^{\circ}$ to $90^{\circ}$ and the SNR is set as 20dB. Other parameters remain the same as in example 1. We configure an optimal transceiver for each scanning angle. The curve of output SINR versus scanning angle is plotted in Fig. \ref{Fig.5}, where we can observe that when the interferences are widely separated from the target, the superiority of the proposed optimal transceiver is not obvious. On the other hand, when the interferences are move closer to the target, the performance of the optimal transceiver exhibits noticeable improvement compared to that of conventional MIMO array.

\begin{figure}[!t]
	\setlength{\abovecaptionskip}{-5pt}
	\setlength{\belowcaptionskip}{0pt}
	\centering
	\includegraphics[height=2.2in, width=3.1in]{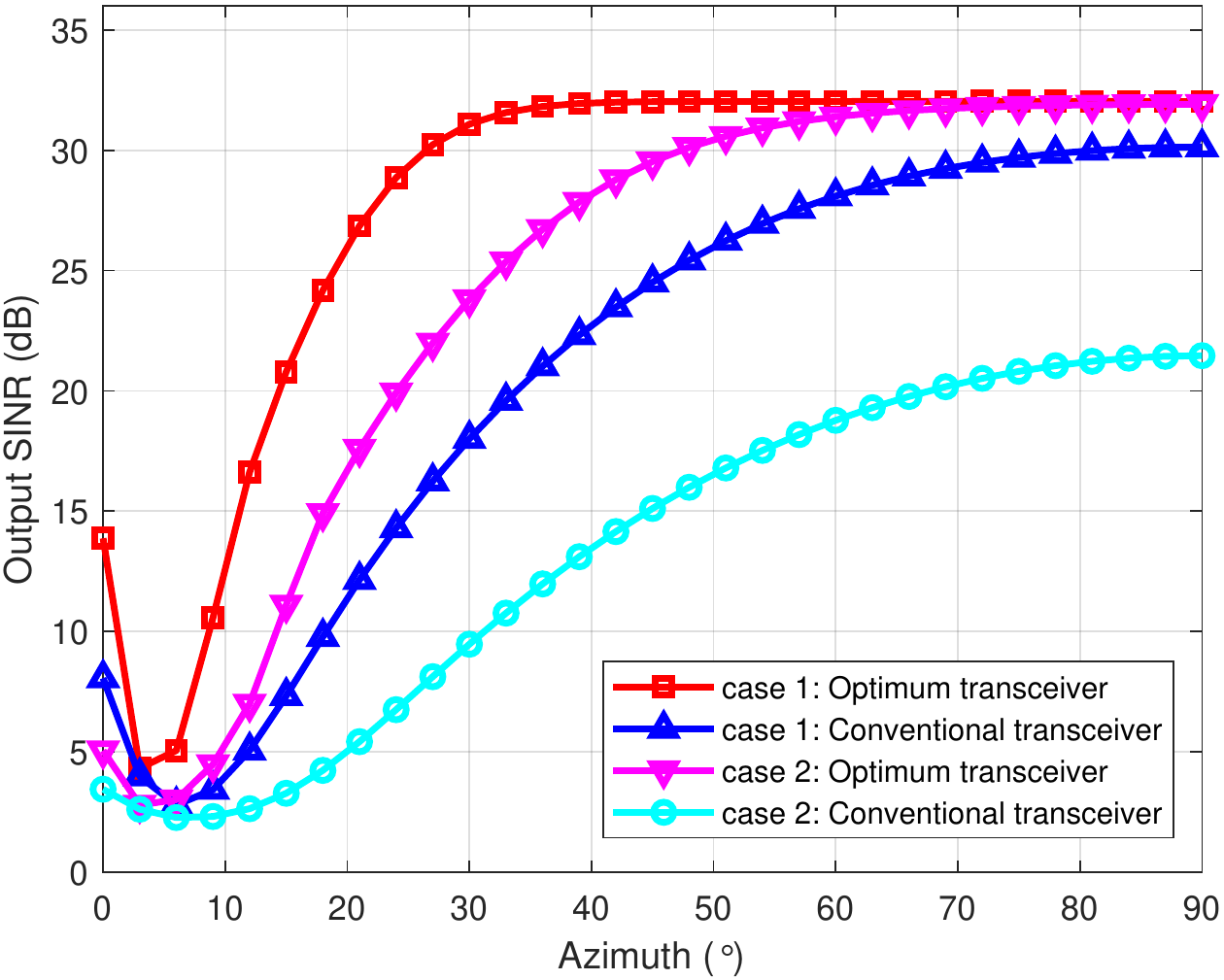}
	\caption{Relationship between the output SINR and the target angle when the interferences are close to the target.}
	\label{Fig.6}
    \vspace{-0.5cm}
\end{figure}
 
\subsection{Example 3}
From example 2, we can argue that the optimal cognitive MIMO radar behaves well when the interferences are close to the target, which is difficult to deal with in practice. Thus, in this example, we focus on this specific scenario. For each scanning angle, we investigate two cases, where the incident angles of two deceptive interference are $\pm 5^{\circ}$ relative to the target in case 1 and then $\pm 3^{\circ}$ in case 2. Again, we plot the curve of output SINR versus the scanning angle, as shown in Fig. \ref{Fig.6}. It can be observed that the optimal transceiver array selected by the optimal cognitive MIMO radar clearly improves the performance. Moreover, the closer the interference is relative to the target, the higher the improvement. Additionally, there is an interesting phenomenon as seen from Fig. \ref{Fig.6}, that the linear MIMO transceiver behaves worse in the endfire direction compared with the broadside direction. The reason is that the beampatterns of linear arrays are broadened in the endfire direction, thus limiting their spatial resolution. 

\subsection{Example 4}
\begin{figure}[!t]
	\setlength{\abovecaptionskip}{0pt}
	\setlength{\belowcaptionskip}{0pt}
	\centering
	\includegraphics[height=2.2in, width=3.1in]{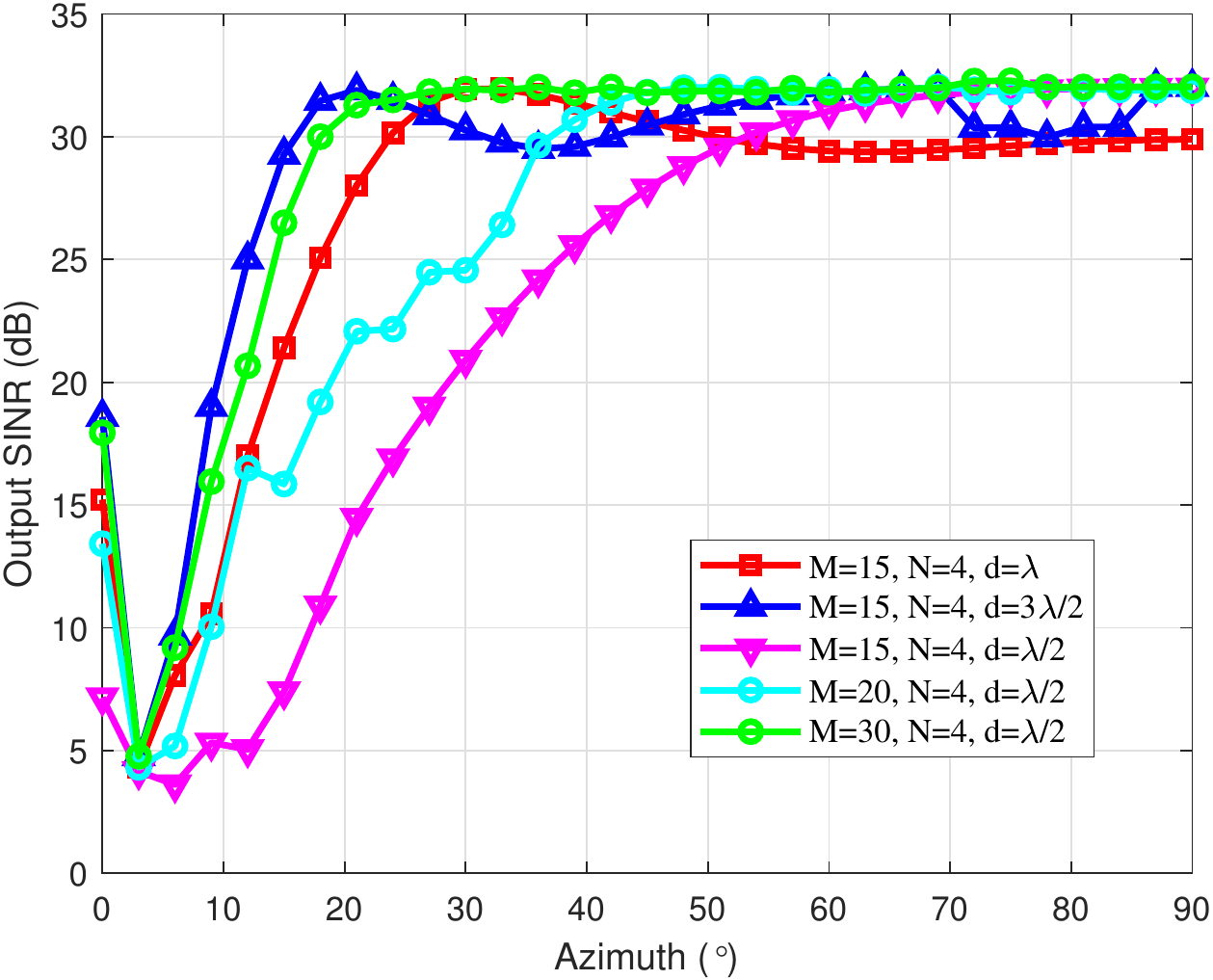}
	\caption{The influence of the total number of sensors and the antenna spacing on the output SINR of optimal sparse transceiver array.}
	\vspace{-0.5cm}
	\label{Fig.7}
\end{figure}

In this example, we examine the effect of the number of candidate antennas and the antenna spacing on the output performance of our optimal sparse transceivers. The output SINR versus the scanning angle is depicted in Fig. \ref{Fig.7}. Again, the linear array exhibits degraded performance when steering in the endfire direction. This phenomenon can be ameliorated by increasing the total array aperture, as proved by the blue curve. We can see that when the sensor spacing is $\lambda/2$, the performance of the optimal sparse transceiver improves with increased number of candidate antennas. For fixed number of antennas, increasing the inter-element spacing improves the performance of the optimal sparse transceiver at lower scanning angle. However, the performance becomes unstable when scanning to broadside direction. 

\section{Conclusion}
\label{conclution}
In this paper, we proposed a novel cognitive MIMO radar which adaptively optimizes both beamforming weights and transceiver array configuration for MaxSINR beamforming. From the same large array, a given number of antennas were selected to construct the transmit and receive arrays jointly. The two arrays were not allowed to have overlapping or adjacent antennas for improved isolation. During perception, we collected the data by sequentially switching among different sets of antennas and constructed the covariance matrix of a full virtual array. This matrix was then incorporated in constrained minimization problem that iteratively minimizes a reweighted mixed $l_{2, 1}$-norm to promote two-dimensional group sparsity for sparse MIMO transceiver design. The simulations showed that the proposed cognitive-driven sparse MIMO array  design can adaptively reconfigure the transceiver to maximize the beamforming performance with dynamically changing environments.
\\
\bibliographystyle{IEEEtran}
\bibliography{references}

\end{document}